\documentclass[AMA,LATO1COL]{WileyNJD-v2}
\usepackage{lineno,hyperref}

\usepackage{multirow}
\usepackage{mathrsfs}
\usepackage{graphicx,epsfig,float}
\usepackage{algpseudocode}
\usepackage{graphics}
\usepackage{epsfig}
\usepackage{amssymb}
\usepackage{algorithm}
\usepackage{algorithmicx}
\usepackage{subfigure}
\usepackage[section]{placeins}
\articletype{Article Type}%

\begin{document}

\title{DE-RSTC: A rational secure two-party computation protocol based on direction entropy}

%\author[myfirstaddress]{}
%\ead{ylwangqfnu@163.com}
%\author[myfirstaddress]{}
%\ead{yangguoyu1020@163.com}
%\author[myfothaddress]{}
%\ead{lifengzhang@ldu.edu.cn}
%\author[myfirstaddress]{}
%\ead{darcy\_wang@163.com}
%\author[mysecondaryaddress]{}
%\ead{kelishan@gzhu.edu.cn}
%\author[myfifthaddress]{}
%\ead{douyi12@gmail.com}
%\author[seventhaddress]{Shouzhe Li}
%\ead{sli649@wisc.edu}
%\author[mysixthaddress]{Xiaomei Yu}
%\ead{yxm0708@126.com}

\author[1,2]{Yuling Chen}
\author[1]{Juan Ma}
\author[3]{Xianmin Wang$^*$}
\author[1]{Xinyu Zhang}
\author[4]{Huiyu Zhou}

\authormark{Yuling Chen \textsc{et al.}}

\address[1]{State Key Laboratory of Public Big Data, College of Computer Science and Technology, Guizhou University, Guiyang, 550025, China}

\address[2]{Guangxi Key Laboratory of Cryptography and Information Security, Guilin University Of Electronic Technology, Guilin, 541000, China}

\address[3]{Institute of Artificial Intelligence and Blockchain, Guangzhou University, Guangzhou, 510000, China}

\address[4]{University of Leicester, Leicester, United Kingdom, Le1 7RH}

\corres{*Xianmin Wang, Institute of Artificial Intelligence and Blockchain, Guangzhou University, Guangzhou, 510000, China. \email{xianmin@gzhu.edu.cn}}

\abstract[Summary]{Rational secure multi-party computation (RSMC) means two or more rational parties complete a function on private inputs. Unfortunately, players sending false information can prevent the protocol from executing correctly, which will destroy the fairness of the protocol. To ensure the fairness of the protocol, the existing works on achieving fairness by specific utility functions. In this paper, we leverage game theory to propose the direction entropy-based solution. To this end, we utilize the direction entropy to examine the player's strategy uncertainty and quantify its strategy from different dimensions. Then, we provide mutual information to construct a new utility for the players. What's more, we measure the mutual information of players to appraise their strategies. By analyzing and proofing of protocol, we show that the protocol reaches a Nash equilibrium when players choose a cooperative strategy. Furthermore, we can solve the fairness of the protocol. Compared to the previous approaches, our protocol is not required deposits and design-specific utility functions. 
}


\maketitle

\section{Introduction}
\label{Introduction}
With the rapid development of big data, the application of cryptography has become more and more popular\cite{36}. In cryptography, secure multi-party computation (SMC) refers to methods for parties to jointly compute a function over their inputs without revealing these values \cite{1,2}. In SMC, the multi-party protocol assumes that parties are honest or malicious. The honest players abide by the rules of the protocol, and the malicious players deviate from the protocol. However, parties make strategic choices to maximize their utility. Therefore, how to ensure the correct implementation of the protocol is an issue that cannot be ignored \cite{2-1,2-2,2-3}. To solve this problem, Halpern \cite{3} first proposed the concept of rational parties, and used rational parties in secret sharing, which played a guiding role in the research of rational secure multi-party computation (RSMC).

RSMC combines game theory and SMC, which is different than SMC. The reason is that it is based on the utility execution protocol. During the execution of the protocol, players have multiple behaviors, which will lead to uncertainty of the protocol. and the uncertainty of events can be measured by information entropy. However, when using information entropy to measure the player's strategy, it cannot measure the positive or negative of the amount of information transmitted. Thus, we increase the positive and negative quantification of information based on information entropy, and we measure the player's strategy from different dimensions. In effect, we use the direction entropy to dynamically quantify the behaviors of the parties to reach the optimal state of the protocol.

Our rational secure two-party computation protocol is a solution based on direction entropy. More concretely, we utilize the direction vector of direction entropy to measure the player's strategy from different dimensions. Next, we utilize mutual information to construct the player's utility function, and analyze the player's interaction in each round to measure the amount of information. Furthermore, we establish a communication channel for the player, and provide the channel capacity to express the maximum utility that the player obtains. Finally, we give a fairness proof of the protocol. Also, we also prove that the protocol has a Nash equilibrium.

\subsection{Related works} %二级标题 相关工作
\label{Related work}
SMC is an emerging topic which has been drawing growing attention during recent decades \cite{5}. The "millionaire" problem proposed by Yao is considered the beginning of secure multi-party computing in 1982 \cite{6}. It is the first one that solved the problem of secure two-party computation. Subsequently. Yao \cite{7} expanded it to solve other problems such as secret sharing. Goldriech et al.\cite{8} extended the two parties to multiple parties and defined the security of secure multi-party computation. So far, there have been many studies on secure multi-party computing based on cryptography \cite{31}. In particular, to solve the fairness in SMC, more research like \cite{27,28,29}, However, in these schemes, although extremely reliable execution can be guaranteed, the cost caused by the cryptographic proof is still expensive.

Game theory is always used to predict the behavior of rational players, where the players are no longer viewed as being honest or corrupt and will not choose a strategy without motivation \cite{10}. Specifically, players who follow the protocol are honest, and players who deviate from the protocol are corrupt. By contrast, all parties are simply rational motivated by some utility function. Since Halpern et al. \cite{3}first proposed the rational player and applied it to the field of secret sharing, lots of research on bridging game theory and cryptography have been done. For instance, wang et al. \cite{11} proposed a rational fair calculation protocol that satisfies a computable sequential equilibrium. Subsequently, they introduced the current research status of RSMC, and discussed the advantages and disadvantages of the previous protocols \cite{12}. In addition, more research on RSMC such as \cite{13,14}.

Since Shannon created information theory, information entropy has been used to measure the uncertainty of events. Since then, lots of research on the uncertainty of measuring events by information entropy have been done \cite{18}. In addition, the researchers studied the fairness \cite{20}, security \cite{17,21}, and privacy \cite{22} issues in secure multi-party computing based on information entropy. In particular, Yang et al \cite{23}. proposed the idea of direction entropy based on information entropy. Specifically, it is based on the information entropy to increase the judgment of the positive and negative of the information. However, existing works guarantee the player's honest strategy by pre-depositing deposits and defining specific utility functions, which is a lack of persuasion since they elaborate designed against set targets.

Recently, there are many kinds of research that use mutual information to study different fields. Tian et al. \cite{24} combined mutual information with rational commissioned calculations to achieve the optimal offensive and defensive strategy of rational commissioned calculations, but only in one-to-one situations. Subsequently, Li et al. \cite{25} extended the one-to-one rational commission calculation to the one-to-many rational delegating computation. In addition, Liu et al. \cite{26} utilized mutual information to obtain the optimal interval of the $k$ value under the $k$-anonymous model. However, in these schemes, the researcher converts the two or more parties' protocol into a communication problem between players. Based on these schemes, we present mutual information to construct a rational player's utility function. Concurrently, when the player's mutual information reaches the channel capacity, its utility is also maximized.

\subsection{Our contribution}
The main contribution of this work is to design a new RSMC protocol by using the direction entropy to measure the player's behavior from different dimensions and redefine the player's utility function. Compared with the existing work, we not required deposits and design-specific utility functions. The fairness of the protocol can be guaranteed. Therefore, compared to the previous solutions, our main contributions are as follows.

\begin{itemize}
	\item Direction entropy. Based on the direction entropy, we propose a direction entropy algorithm for rational secure two-party computation. More specifically, we provide the direction vector to investigate the player's strategy uncertainty from different dimensions. If the player chooses the cooperation strategy, and its direction vector is positive, which means that the information transmitted by the player is positive and vice versa. 

	\item New utility function. Perhaps more interestingly, we analyze the player's behavioral strategies, and we employ mutual information to represent the player's new utility function. Next, when the information transmitted by the player is positive, the number of mutual information increases. In other words, if the player chooses a cooperative strategy, its utility is raised. Conversely, the utility is reduced. Then, we construct a communication channel for the player, and measure the player's channel capacity to determine the player's behavior strategy.

	\item Achieve fairness. We prove in detail the protocol we gave, and the result show that the fairness problem in rational secure two-party computation. We prove that the proposed protocol reaches the Nash equilibrium. Moreover, we analyze our entire scheme and conclude that the protocol we design maximizes utility for players. In a nutshell, the protocol reaches the Nash equilibrium when the probability of players choosing to cooperate approaches 1.
\end{itemize}

\subsection{Roadmap} %二级标题 

The rest of the paper is structured as follows. Section \ref{Introduction} presents the related work and our contribution. Section \ref{Preliminaries} introduces concepts such as game theory, information entropy, direction entropy, mutual information, and channel capacity. Section \ref{sec:Information exchange protocol based on direction entropy.} proposes an information exchange protocol based on direction entropy. Section \ref{sec:New utility based on mutual information.} constructs the player's utility function based on mutual information. Section \ref{Proof} proves and analyses the protocol. Section \ref{sec: Simulations and results}  simulates the player's utility and analyzes the performance of the protocol. The paper concludes with a summary and future directions in Section \ref{sec:Conclusion}.

%第二章
\section{PRELIMINARIES} 
\label{Preliminaries}
In this section, we introduce the basic knowledge needed for our scheme. Firstly, we introduce the concept of the standard game. Furthermore, we review the most important concept in game theory, i.e., the Nash equilibrium. Then, we give the definition of information entropy and direction entropy. Finally, we introduce some basic knowledge in information theory. %一级标题 相关知识

\subsection{Game theory}
\textbf{Definition 1 (Standard Game).} The standard form of a $n$-player game is composed of three elements: player set $P$, strategy space $S$ and utility function $u$, denoted as $G=\{P,S,u\}$, where $P=\{P_{1},\cdots,P_{n}\}$, $S=\{S_{1},\cdots,S_{n}\}$, $u=\{u_{1},\cdots,u_{n}\}$. Any specific strategy $s_{i} \in S_{i}$ indicates that strategy $s_i$ is the key element of the strategy set $S_i$, and utility function $u: S \rightarrow R$ denotes the profits of the players $i$ under different strategy profiles.

\textbf{Definition 2 (Nash Equilibrium).} A strategy profile $s^{*}=\{s_{1}^{*},\cdots,s_{n}^{*}\}$ is a Nash equilibrium of game  $G=\{P,S,u\}$, if $u_{i}(s_{i}^{*},s_{-i}^{*}) \ge u_{j}(s_{j}^{*},s_{-j}^{*})$ holds for each player $P_{i} (i=1,\cdots,n)$ and all $s_{j} \in S_{i}$. 

Obviously, if player $i \ne j$ complies with the strategy $s_{i}^{*}$, then the player will not deviate from the strategy $s_{j}^{*}$, as it will not benefit at all. In principle, there may be multiple Nash equilibrium in a game.

\subsection{Entropy}
\textbf{Definition 3 (Information Entropy).} Information entropy is a measure of the degree of uncertainty. If a thing  $X$ has $n$ mutually exclusive possible outcomes, and the probability of occurrence of the $n$ outcomes is $\{p_{1},\cdots,p_{n}\}$  respectively, the entropy to measure the degree of uncertainty of the thing $X$ is defined as:
\begin{equation}
	H(X)=\sum_{i=1}^np_ilog\dfrac{1}{p_i}
\end{equation}

\textbf{Definition 4 (Direction Entropy).} 
Direction entropy is based on the information entropy to increase the measurement of the positive and negative information. Specifically, it refers to a random variable $X$ composed of n random events with probabilities $\{p_{1},\cdots,p_{n}, p_{1}+,\cdots,+p_{n}=1\}$, and the total amount of information contained is:
\begin{equation}
	H(\textbf{a},X)=\sum_{i=1}^na_ip_ilog\dfrac{1}{p_i}
\end{equation}

\subsection{Information Theory}\label{sec1}
\textbf{Definition 5 (Mutual Information).} Another random variable causes to reduce the decrement of uncertainty of the original random variable. The mutual information denotes the decrement of uncertainty about $X$ between before and after receiving $Y$. It is also the metric of independent degree between two random variables. It is symmetrical about $X$ and $Y$ and non-negative, the mutual information is 0 if and only if $X$ and $Y$ are mutually independent. 
\begin{equation}
	H(X;Y)=H(X)-H(X\|Y)=\sum_{x,y}p(x,y)log\dfrac{p(x,y)}{p(x)p(y)}
\end{equation}

\textbf{Definition 6 (Channel Capacity).} 
\label{Definition 6 (Channel Capacity).}
For the communication channel where the input signal is  $X$ and $Y$ the output signal is $Y$, the channel capacity $C$ defined as  
\begin{equation}
	C=\underset{p(x)}{max}\ I(X;Y)
\end{equation}

\section{Information exchange protocol based on direction entropy}
\label{sec:Information exchange protocol based on direction entropy.}
In this section, we design a new protocol to measure rational player strategy based on direction entropy. First, we examine the player's strategy uncertainty from different dimensions. Moreover, we use the direction vector of the direction entropy to quantify the positive and negative of the information transmitted by the player.

\subsection{Rational secure two-party computation framework}
In this section, we first describe the parameters and concepts required in this solution. Specifically, we use the following notations in Table \ref{tab:Notations.}.

\begin{table}[htbp]
		\centering
	\caption{Notations.}
	\label{tab:Notations.}
	\begin{tabular}{ccc}
	\toprule  % 中部线
		Serial&Symbol & Description \\
		\toprule
		(1)&$\mu$& $P_1$ probability of choosing cooperation \\
		(2)&$\nu$& $P_2$ probability of choosing cooperation \\
		(3)&$C$& The cooperative strategy \\
		(4)&$D$& The non-cooperative strategy \\
		(5)&$S$& The secret\\
		(6)&$H(dir_{t})$& Directional entropy at round $t$\\
		(7)&$H(dir_{t-1})$& Directional entropy at round $t-1$\\
		(8)&$\textbf{A}\in\{C,D\}$& The strategy set\\
		(9)&$s_{1}^1,\cdots,s_{1}^m$& Sub-secret of player $P_1$ \\
		(10)&$s_{2}^1,\cdots,s_{2}^m$& Sub-secret of player $P_2$\\
		(11)&$P_{i}\in \textbf{P}(i\in\{1,2\})$& The party set\\
		\bottomrule
	\end{tabular}
\end{table}

As illustrated in the framework of the protocol, our protocol consists of three main phases described in detail in the following.

Phase 1: At this stage, the initialization of the protocol is mainly completed (Initialization of direction entropy is given in section \ref{sec:Initialization of direction entropy.}).

Phase 2: In the second phase, this phase completes the information exchange of the protocol (Information exchange based on direction entropy is given in section \ref{sec:Information exchange based on direction entropy}).

Phase 3: At this stage, this phase completes the information computation of the protocol  (Computation based on direction entropy is given in section \ref{sec:Compute the player's direction entropy}).

Phase 4: Finally, the integration phase is where the participant integrates all received sub-secrets to recover the secret.

Three main phases of the proposed protocol are summarized in the following.

\begin{center}
	\fbox{
		\begin{minipage}[h][][c]{13cm}
			\begin{center}
				\textbf{\textit{protocol $\pi$ }}
			\end{center}
			\textbf{Phase 1: Initialization phase}\\
			Step 1: TTP divides the $S$ into $m$ shares, ie, $s_{1}^1,\cdots,s_{1}^m$ and $s_{2}^1,\cdots,s_{2}^m$.\\
			Step 2: TTP sends the $s_{1}^1,\cdots,s_{1}^m$ to $P_1$ and $s_{2}^1,\cdots,s_{2}^m$ to $P_2$.\\
			\textbf{Phase 2: Exchange phase}\\
			During the $t$-th round of secret exchange, there are the following steps:\\
			Step 3: $P_1$ sends sub-secret $s_{1}^t$ to $P_2$.\\
			Step 3.1: If $P_1$ choose $C$, then $a_t=1$, there is $\Delta H_{P_1}(dir>0)$;\\
			Step 3.2: Otherwise, $a_t=-1$, there is $\Delta H_{P_1}(dir\le0)$.\\
			Step 4: $P_2$ sends sub-secret $s_{2}^t$ to $P_1$.\\
			Step 4.1: If $P_2$ choose $C$, then $b_t=1$, there is $\Delta H_{P_2}(dir>0)$;\\
			Step 4.2: Otherwise, $b_t=-1$, there is $\Delta H_{P_2}(dir\le0)$.\\
			\textbf{Phase 3: Computation phase}\\
			Step 5: Once $P_i$ receives the sub-secret from $\overline{P_i}$, we use the public function $f$ to calculate: $f_i(s_{i}^t)\to y_i(I(S;s_{i}^t))$. We use $s_j$ to denote the sub-secrets that have been successfully exchanged. The specific steps are as follows:\\
			Step 5.1: If $I(S;s_{i}^t)\neq I(S;s_{j})$, then $P_i$ successfully obtains a part of $\overline{P_i}$'s secret;\\
			Step 5.2: If $I(S;s_{i}^t)=I(S;s_{j})$, then $P_i$ obtains a duplicate sub-secret of $\overline{P_i}$;\\
			Step 5.3: If $I(S;s_{i}^t)=0$, then $P_i$ obtains  an incorrect sub-secret of $\overline{P_i}$.\\
			\textbf{Phase 4: Integration phase}\\
			When $n$ rounds of protocol interaction is completed, if no player terminates the protocol during the execution of the protocol, the result of the protocol is that both players obtain all correct sub-secrets, and the secret is restored by integrating all sub-secrets, which is also called the correct result value. Subsequently, both players withdrew from the protocol.
		\end{minipage}
	}
\end{center}

\subsection{Initialization of direction entropy}
\label{sec:Initialization of direction entropy.}
As mentioned definition 4,  $n$ is the number of factors in the agreement that are all independent of each other and affect the player's strategy, and $p_i$ is the probability of an event triggered by the $i$-th factor that affects the player's strategy. This includes all the factors that affect the player's strategy, so there is a probability normalization condition:
\begin{equation}
	\sum_{i=1}^na_ip_i=1 
\end{equation}

We assume that the factors that affect the strategy in the agreement are $p_{1},\cdots,p_{n}$ (such as complexity; utility; time, etc.), and the direction entropy at round $t$ is $H(dir_{t})$. Meanwhile, $H(dir_{t})$ increases with the number of rounds, that is, $\dfrac{dH(dir_{t})}{d_t}>0$. It means that the player complied with the protocol during the t round. Conversely, $H(dit_{t})$ decreases with the number of rounds, that is, $\dfrac{dH(dir_{t})}{d_t}<0$. It indicates that the player deviated from the protocol during the $t$ round. 

(1) First, we assume that a trusted third-party TTP divides the secret $S$ into $m$ shares, and sends the $S$ shares to the player $P_i\in\textbf{P}(i\in{1,2})$.

(2) TTP chooses $s_{1}^1,\cdots,s_{1}^m$ and $s_{2}^1,\cdots,s_{2}^m$ as the exchange secret $S$, and they are held by $P_1$ and $P_2$ respectively.

(3) $P_1$ and $P_2$ exchange secrets $s_{1}^1,\cdots,s_{1}^m$ and $s_{2}^1,\cdots,s_{2}^m$  for $n$ $(m<n)$ rounds, and finally $P_1$ and $P_2$ integrate the $m$ secrets received to form $S$. 

In the exchange phase, player $P_i\in\textbf{P}(i\in{1,2})$ has two behavior strategies, namely cooperative strategy $C$ or non-cooperative strategy $D$. The strategy set is $\textbf{A}\in\left\{C,D\right\}$ . We combine the strategies of the players to get the strategy combination matrix. The details are as follows:

$$\textbf{A}_{2\times 2}=\begin{bmatrix}
	(C,C) & (C,D) \\
	
	(D,C) & (D,D) \\
\end{bmatrix}$$

The above strategy matrix has the following four situations:

Case 1: $(C,C)$ means that both  $P_1$ and $P_2$ choose a cooperation strategy.

Case 2: $(C,D)$ means that the $P_1$ chooses a cooperative strategy, while  $P_2$ chooses a non-cooperative strategy.

Case 3: $(D,C)$ means that the $P_1$ chooses a non-cooperative strategy, while $P_2$ chooses a cooperative strategy.

Case 4: $(D,D)$ means that both $P_1$ and $P_2$ choose a non-cooperative strategy.

In addition, there are two cases for the direction vectors $\textbf{a}=(a_1,a_2,\cdots,a_n)$ and $\textbf{b}=(b_1,b_2,\cdots,b_n)$ of the players $P_1$ and $P_2$. Specifically, when the player chooses the cooperative strategy $C$, the direction vector is positive, and when the player chooses the non-cooperative strategy $D$, the direction vector is negative. We have the following assumptions:

(1) The probability that the $P_1$ chooses $C$ is $\mu$, its direction vector is positive, the probability of choosing $D$ is $1-\mu$, its direction vector is negative, and satisfies $\mu+(1-\mu)=1$.

(2) The probability that the $P_2$ chooses $C$ is $\nu$, its direction vector is positive, the probability of choosing  $D$ is $1-\nu$, its direction vector is negative, and satisfies $\nu+(1-\nu)=1$.

\subsection{Information exchange based on direction entropy}
\label{sec:Information exchange based on direction entropy}
In the information exchange process in the secure two-party computation, a total of $n$ rounds of information needs to be exchanged. In each round of information exchange, it is assumed that the player has a current round direction entropy $H_{P_i}(dir_{cur})$. We use direction entropy to evaluate participants' strategies. In $t\in\left\{1,2,\cdots,n\right\}$ rounds. As follows:

Phase 1: Player  $P_i\in\textbf{P}(i\in{1,2})$ (assuming player $P_1$ ) chooses a strategy from strategy set $\textbf{A}$. If  chooses the cooperation strategy $C$, then $P_1$ will send $s_1^t$ to $P_2$. In other words, the direction entropy of $P_1$ is $H(dir_t)$ at round $t$. If chooses the non-cooperative strategy $D$, then $P_1$ will not send $s_i^t$ to $P_2$, the protocol is terminated, and $P_2$ can only choose $D$. That is, neither $P_1$ nor $P_2$ sends a secret, and their direction entropy is $H(dir_{t-1})$.

Phase 2: If $P_1$ chooses strategy $C$ in the first stage, $P_2$ chooses a strategy from strategy set $\textbf{A}$. If $P_2$ chooses the cooperation strategy $C$, then $P_2$ will $s_2^t$ send  to $P_1$. In other words, the direction entropy of $P_1$ and $P_2$ are $H(dir_t)$. If $P_2$ chooses the non-cooperative strategy $D$, then $P_2$ will not send $s_2^t$ to $P_1$, and the protocol is terminated, that is, the direction entropy $P_2$ of is $H(dir_{t-1})$. If no participant adopts a non-cooperative strategy $D$ and $t<n$, the protocol enters the next round. Otherwise, the protocol terminates.

Specifically, in round  $t\in\left\{1,2,\cdots,n\right\}$, the following four situations occur:

Case 1: If $P_1$ chooses a cooperative strategy, the direction vector of its direction entropy is positive. At the same time, events with probability $\mu$ provide positive information with a quantity of ${\mu}log\dfrac{1}{\mu}$ because $P_1$ chooses a cooperative strategy with probability $\mu$. In short, in the $t$-th round, the direction value of $P_1$ in the $t$-th round is greater than the direction deletion value in the $t-1$ round,  i.e.,$H(dir_t)> H(dir_{t-1})$.

Case 2: If $P_1$ chooses a non-cooperative strategy, the direction vector of its direction entropy is negative. At the same time, events with probability $1-\mu$ provide positive information with a quantity of ${-(1-\mu)}log\dfrac{1}{(1-\mu)}$ because $P_1$ chooses a cooperative strategy with probability $1-\mu$. In short, in the $t$-th round, the direction value of $P_1$ in the $t$-th round is less than the direction deletion value in the $t-1$ round,  i.e.,$H(dir_t)\le H(dir_{t-1})$. 

Case 3: If $P_2$ chooses a cooperative strategy, the direction vector of its direction entropy is positive. At the same time, events with probability $\nu$ provide positive information with a quantity of ${\nu}log\dfrac{1}{\nu}$ because $P_2$ chooses a cooperative strategy with probability $\nu$. In short, in the $t$-th round, the direction value of $P_2$ in the $t$-th round is greater than the direction deletion value in the $t-1$ round,  i.e.,$H(dir_t)> H(dir_{t-1})$.

Case 4: If $P_2$ chooses a non-cooperative strategy, the direction vector of its direction entropy is negative. At the same time, events with probability $1-\nu$ provide positive information with a quantity of ${-(1-{\nu})}log\dfrac{1}{(1-\nu)}$ because $P_2$ chooses a cooperative strategy with probability $1-\nu$. In short, in the $t$-th round, the direction value of $P_2$ in the $t$-th round is less than the direction deletion value in the $t-1$ round,  i.e.,$H(dir_t)\le H(dir_{t-1})$.  

For the proposed protocol, we propose an algorithm to measure the player's behavioral uncertainty and changes in direction entropy. The specific algorithm details are shown in \textbf{Algorithm} \ref{Algorithm 1}:
\renewcommand{\algorithmicrequire}{\textbf{Input:}}
\renewcommand{\algorithmicensure}{\textbf{Output:}}
\begin{algorithm}[h]
	\caption{The change of directional entropy.} 
	\label{Algorithm 1}
\begin{algorithmic}[1]
		\Require{$t$, $H(dir_{t-1})$.}
	\Ensure{$H(dir_t)$.}
		%初始化
	\State 	Initialize $s_{1}^{1},...,s_{1}^{m} \in random (S_1)$ 
	\State 	Initialize $s_{2}^{1},...,s_{2}^{m} \in random(S_2) $ 
	\State 	Initialize $\textbf{A}$=$\{ C,D\}$.
	\State 	Initialize $t\leftarrow1$.
		\For  {t $<$ n}{
			\State	$P_{2}\leftarrow P_{1}$ $chooses$ sub-secret $s_{1}^{t}$.
				\If {$s_{1}^{t}$$\in$$ random (S_1)$} 
				\State	$a_{i} =1$ ;
					\State$H_{P_1}(dir_t)>H_{P_1}(dir_{t-1})$;
				\Else
					\State	$a_{i} =-1$;
					\State	$H_{P_1}(dir_t)\le H_{P_1}(dir_{t-1})$.
					\State $t \leftarrow t+1$.
				\EndIf
			\State	$P_{1}\leftarrow P_{2}$ $chooses$ sub-secret $s_{2}^{t}$.
			\If {$s_{2}^{t}$$\in$$ random (S_2)$} 
				\State	$b_{i} =1$ ;
					\State$H_{P_2}(dir_t)>H_{P_2}(dir_{t-1})$;
				\Else
				\State	$b_{i} =-1$; 
						\State$H_{P_2}(dir_t)\le H_{P_2}(dir_{t-1})$.
					\State$t \leftarrow t+1$.
				\EndIf
			\EndFor 
		}
\end{algorithmic}
\end{algorithm}

According to the above \textbf{Algorithm} \ref{Algorithm 1}, we know that players have different strategies. For players' different strategies, we use direction entropy to measure players' strategies in multiple dimensions. Specifically, when the players cooperate, the player's direction vector is positive. Otherwise, the player's direction vector is negative.

\subsection{Computation of the player's direction entropy}
\label{sec:Compute the player's direction entropy}
After the end of round $t$, we calculate the direction entropy difference of player $P_i$. We use the direction entropy difference to indicate the change of the direction entropy value of the players after a round of exchange of information, which can help the players to understand each other better in the subsequent exchange process. We assume that the player's direction entropy at round $t-1$ is $H(dir_{t-1})$, and after the $t$-th round of exchange, $P_i$'s direction entropy is $H(dir_{t})$. There are the following computations:
\begin{equation}
	\label{equ6}
	\Delta H_{P_i}(dir)=H_{P_i}(dir_t)+H_{P_i}(dir_t-1)
\end{equation}

According to formula \ref{equ6} above,ru1suanfa1 we get the following two cases:

Case 1: $\Delta H_{P_i}(dir>0)$ indicates that the direction entropy of $P_i$ in the $t$-th round is greater than the direction entropy of the $t-1$ round.

Case 2: $\Delta H_{P_i}(dir\le0)$ means that the direction entropy of  in the $t$-th round is less than the direction entropy of the $t-1$ round.

After the $t$-th exchange, the direction entropy of $P_i$ is expressed as follows:
\begin{equation}
	\Delta 	H_{P_i}(dir_t)=\left\{
	\begin{array}{ll}
		H_{P_i}(dir_{t-1}),       P_i\in $D$\\
		H_{P_i}(dir_{t-1})+H_{P_i}(dir_{cur}),  P_i\in $C$
	\end{array}
	\right.
\end{equation}

Additionally, we give a detailed algorithm for information exchange based on direction entropy, which is described as follows:
\renewcommand{\algorithmicrequire}{\textbf{Input:}}
\renewcommand{\algorithmicensure}{\textbf{Output:}}
\begin{algorithm}[h!]
	\caption{Computation of direction entropy} \label{Algorithm 2}
\begin{algorithmic}[1]
		\Require{$H_{P_{i}}(dir_{t-1})$, $H_{P_{i}}(dir_{t})$. }
		\Ensure{$\Delta H_{P_{i}}(dir)$.}
	\State 	Initialize  $t\leftarrow1$.
	\State 	Initialize  $P_{i}(i\in1,2)$.
\State $P_{i}$ Compute $\Delta H_{P_{i}}(dir)=H_{P_{i}}(dir_t)-H_{P_{i}}(dir_{t-1}).$
		%初始化
		\For{$t<n$}
			\If  {$\Delta H_{P_{i}}(dir) > 0$} 
				\State	$P_{i}$ sends $s_{i}^{t}$;
					\State$t \leftarrow t+1$;
				\Else 
					\State	$P_{i}$ refuses to send $s_{i}^{t}$;
					\State	Protocol termination. 
				\EndIf
			\EndFor
\end{algorithmic}
\end{algorithm}

As can be seen from \textbf{Algorithm} \ref{Algorithm 2}, each player quantifies the direction entropy of the opposing player in round $t$. Players are more inclined to trust the opposing participants with higher direction entropy to choose cooperation strategies to exchange information. For players with low direction entropy, they are more inclined to distrust that they will choose a cooperative strategy to exchange information. 

\section{New utility based on mutual information}
\label{sec:New utility based on mutual information.}
In this part, we construct utility functions for rational players based on mutual information. First, we define the player's utility function, which is mutual information. Then, we construct a standard game and communication channel for the player. Finally, we calculate the mutual information of each player in detail.
\subsection{New definition of utility}\label{subsec3}

In this section, we first provide a standard game theory model $G=\{P_1,P_2,S_1,S_2;u_1,u_2\}$, where the direction vectors of players $P_1$ and $P_2$ are $\textbf{a}=(a_1,a_2, ... ,a_n)$ and $\textbf{b}=(b_1,b_2, ... ,b_n)$ respectively. Next, we assume that a transition matrix $\textbf{A}=[A_{ij}]$, which is equivalent to determining the transition matrix of a certain channel, namely $A_{ij}=p(x=i \vert y=j), 1 \leq i\leq n,1 \leq j \leq m$. If $P_1$ and $P_2$ are respectively $n$ and $m$ random variables. The strategic space $S_1$ of $P_1$ is defined as $S_1=\{0 \leq x_i \leq 1; 1 \leq i \leq n,x_1+x_2+,...,+x_n=1\}$, and the strategic space $S_2$ of $P_2$ is defined as $S_2=\{0 \leq y_i \leq 1; 1 \leq i \leq m,y_1+y_2+...+y_m=1\}$. $P_1$ and $P_2$ are arbitrary pure strategies any $s_1 \in S_1 (ie.,s1=(p_1,p_2,...,p_n),p_1+p_2+...+p_n)$ and $s_2 \in S_2 (ie.,s_2=(q_1,q_2,...,q_n),q_1+q_2+...+q_n)$. In this setting, the utility function $u_1(s_1,s_2)$ of player   is defined as:
\begin{equation}
	u_1(s_1,s_2)= \sum_{j=1}^{m}q_j \sum_{i=1}^{n}A_{ij} \log\frac{A_{ij}}{p_i}
\end{equation}

Similarly, the utility function $u_2(s_1,s_2)$ of player $P_2$ is defined as:
\begin{equation}
	u_2(s_1,s_2)= \sum_{i=1}^{n}p_i \sum_{j=1}^{m}B_{ji}\log \frac{B_{ij}}{q_j}
\end{equation}

In summary, the utility function of player $P_1$ is $u_1(s_1,s_2)$, i.e., $I(X;Y)$. Similarly, the utility function of $P_2$ is $u_2(s_1,s_2)$ of $P_2$ is $I(Y;X)$. Where the probability distribution functions of $X$ and $Y$ are defined by $s_1$ and $s_2$ as: $P(X=i)=p_i,(1 \leq i \leq n,0 \leq p_i \leq 1)$ and $P(Y=j)=q_j,(1\leq j \leq m,0 \leq q_j \leq 1)$. 

Among the operations, we use mutual information to represent the player's utility function. Next, we provide direction entropy to measure the player's behavior uncertainty. Players choose different strategies, the direction vector of the direction entropy is different, and the amount of positive and negative information transmitted is also different. Therefore, the player's utility function is defined separately based on the direction entropy. 

In the following, we give mutual information to represent the player's new utility function described in detail.

(1) The utility function $u_1(s_1,s_2)$ of the player $P_1$ is defined as $u_1(s_1,s_2)=\vert I(\textbf{a},X;\textbf{b},Y)\vert $, which is the absolute value of the direction mutual information $\vert I(\textbf{a},X;\textbf{b},Y) \vert$  of $X$ and $Y$.

(2) The utility function $u_2(s_1,s_2)$ of the player $P_2$ is defined as $u_2(s_1,s_2)=\vert I(\textbf{b},Y;\textbf{a},X)\vert$, which is the absolute value of the direction mutual information $\vert I(\textbf{b},Y;\textbf{a},X)\vert $ of $Y$ and $X$.

\subsection{New utility function of the players}\label{subsec3}
In this section, We first use $X=0$ to represent the players $P_1$ chooses a non-cooperative strategy, and $X=1$ to represent the players $P_1$ chooses a cooperative strategy. Similarly, We use $Y=0$ to represent the players $P_2$ chooses a non-cooperative strategy, and $Y=1$ to represent the players $P_2$ chooses a cooperative strategy. Then the behavior strategies of $P_1$ and $P_2$ can be described by the probability distribution of $X$ and $Y$ respectively, as shown in Equation \ref{equ10}:
\begin{equation}
	\label{equ10}
	\begin{aligned}
		&P_r(X=0)=1-\mu;P_r(X=1)=\mu\\
		&P_r(Y=0)=1-\nu;P_r(Y=1)=\nu
	\end{aligned}
\end{equation}

Where $0<\mu<1$, $0<\nu<1$. According to the probability distribution of $P_1$ and $P_2$, we suppose the joint probability distribution of random variable $(X,Y)$ is as follows:
\begin{equation}
	\begin{aligned}
		&P_r(X=0,Y=0)=a\\
		&P_r(X=0,Y=1)=b\\
		&P_r(X=1,Y=0)=c\\
		&P_r(X=1,Y=1)=d\\
		&0<a,b,c,d,\mu , v<1
	\end{aligned}
\end{equation}

Where $a,b,c,d,\mu,\nu$ satisfy the following three linear relationship equations:
\begin{equation}
	\begin{split}
		&a+b+c+d=1\\
		&\mu =P_r(X=0,Y=0)+P_r(X=0,Y=0)=a+b\\
		&\nu =P_r(X=0,Y=0)+P_r(X=1,Y=0)=a+c
	\end{split}
\end{equation}

Next, we construct a communication channel $(X;Y)$ for the player $P_1$ with $X$ as input and $Y$ as output. Specifically, we have the following theorem:

\textbf{Theorem 1}. We assume the channel capacity  $M$ of the  player $p_1$ channel composed of a random variable $(X;Y)$.

(1) If the player wants to obtain the sub-secret share of the $t$-th round, then there must be some kind of skill that enables it to achieve the goal with a probability close to 1 during the $t/M$ exchange.\par
(2) If the player wins $S$ times during $n$ rounds of exchange, then there must be $S\leq nM$.\par
From the above theorem, we only require the channel capacity $M$ of the player's $P_1$ channel, then the player's utility limit is determined, that is, the maximum utility function of $P_1$. Its $2 \times 2$ order transition probability matrix $\textbf{A}=[A(i,j)],i,j=0,1$, the details are as follows:
\begin{equation}
	\begin{split}
		&A(0,0)=P_r(Y=0 \vert X=0)=\frac{P_r(Y=0, X=0)}{P_r(X=0)}=\frac{a}{\mu}\\
		&A(0,1)=P_r(Y=1 \vert X=0)=\frac{P_r(Y=1,X=0)}{P_r(X=0)}=\frac{b}{\mu}=1-\frac{a}{\mu}\\
		&A(1,0)=P_r(Y=0 \vert X=1)=\frac{P_r(Y=0,X=1)}{P_r(X=1)}=\frac{c}{1-\mu}=\frac{\nu-a}{1-\mu}\\
		&A(1,1)=P_r(Y=1 \vert X=1)=\frac{P_r(Y=1, X=1)}{P_r(X=1)}=\frac{d}{1-\nu}=1-\frac{\nu-a}{1-\mu}
	\end{split}
	\nonumber
\end{equation}

Then, we get the transfer matrix of $P_1$ is as follows:

$$A_{2 \times 2}=\begin{bmatrix}
	A(0,0)& A(0,1)\\ 
	A(1,0)&	A(1,1) 
\end{bmatrix}=\begin{bmatrix}
	\frac{a}{\mu}& 1-\frac{a}{\mu}\\ 
	\frac{\nu-a}{1-\mu}& 1-\frac{\nu-a}{1-\mu}
\end{bmatrix}$$

According to the transition probability matrix of player $P_1$, we define the utility function $u_1(s_1,s_2)$ of $P_1$ as the following formula:
\begin{equation}
\label{equation 13}
	\begin{split}
		&u_1(s_1,s_2)=\vert I(\textbf{a},X;\textbf{b},Y)\vert\\
		&= \vert\sum_{\textbf{a},x}  \sum_{\textbf{b},y}p(x,y)\log\frac{p(x,y)}{p(x)p(y)}  \vert \\
		&= \vert  a\log\frac{a}{{\mu}\nu}-b\log\frac{b}{\mu(1-\nu)}-c\log\frac{c}{\nu(1-\mu)}-d\log\frac{d}{(1-\mu)(1-\nu)}  \vert \\
		&= \vert  a\log\frac{a}{{\mu}\nu}-(\mu-a)\log\frac{\mu-a}{\mu(1-\nu)}-(\nu-a)\log\frac{\nu-a}{\nu(1-\mu)}-(1+a-\mu-\nu)\log\frac{1+a-\mu-\nu}{(1-\mu)(1-\nu)}\vert  
	\end{split}
\end{equation}

Next, similar to \textbf{Theorem 1}, we construct a communication channel $(Y;X)$ for the player $P_2$ with $Y$ as input and $X$ as output. To this end, we have the following \textbf{Theorem 2}.

\textbf{Theorem 2}. We assume the channel capacity $N$ of the player $P_2$ channel composed of a random variable $(Y;X)$.

(1) If the player wants to obtain the sub-secret share of the $t$-th round, then there must be some kind of skill that enables it to achieve the goal with a probability close to 1 during the $t/N$ exchange.

(2) If the player wins $S$ times during $n$ rounds of exchange, then there must be $S \leq nN$.

From the above theorem, we only require the channel capacity $N$ of the player's $P_2$ channel, then the player's utility limit is determined, that is, the maximum utility function of $P_2$. The transition probability matrix $\textbf{B}=[B(i,j)],i,j=0,1$, the details are as follows:
\begin{flalign}
	\begin{split}
		&B(0,0)=P_r(X=0 \vert Y=0)=\frac{P_r(X=0,Y=0)}{P_r(Y=0)}=\frac{a}{\nu}\\
		&B(0,1)=P_r(X=1 \vert Y=0)=\frac{P_r(X=1,Y=0)}{P_r(Y=0)}=\frac{c}{1-\nu}=\frac{\nu-a}{1-\nu}\\
		&B(1,0)=P_r(X=0 \vert Y=1)=\frac{P_r(X=0, Y=1)}{P_r(Y=1)}=	\frac{b}{\nu}=1-\frac{a}{\nu}\\
		&B(1,1)=P_r(X=1 \vert =1)=\frac{P_r(X=1, Y=1)}{P_r(Y=1)}=\frac{d}{1-\nu}=1-\frac{\nu-a}{1-\nu}
	\end{split}
	\nonumber 
\end{flalign}

Then, similar to $P_1$, we get the transfer matrix of $P_2$ is as follows:

$$B_{2 \times 2}=\begin{bmatrix}
	B(0,0)& B(0,1)\\ 
	B(1,0)&	B(1,1) 
\end{bmatrix}=\begin{bmatrix}
	\frac{a}{\nu}& 1-\frac{a}{\nu}\\ 
	\frac{\nu-a}{1-\nu}&1-\frac{\nu-a}{1-\nu}
\end{bmatrix}$$

Through the transition matrix we obtained above, we show the utility function $u_2(s_1,s_2)$ of $P_2$:
\begin{equation}
	\begin{split}
		&u_2(s_1,s_2)=\vert I(\textbf{b},X;\textbf{a},Y) \vert\\
		&=\vert\sum_{\textbf{b},y}  \sum_{\textbf{a},x}p(x,y)\log\frac{p(y,x)}{p(y)p(x)} \vert \\
		&=\vert a\log\frac{a}{{\mu}\nu}-b\log\frac{b}{\nu(1-\mu)}-c\log\frac{c}{\mu(1-\nu)}-d\log\frac{d}{(1-\mu)(1-\nu)} \vert \\
		&=\vert a\log\frac{a}{\mu \nu}-(\mu-a)\log\frac{\mu-a}{\mu(1-\nu)}-(\nu-a)\log\frac{\nu-a}{\nu(1-\mu)}-(1+a-\mu-\nu)\log\frac{1+a-\mu-\nu}{(1-\mu)(1-\nu)} \vert
	\end{split}
\end{equation}

\section{Proof}
\label{Proof}
At first, we prove that the player realizes the Nash equilibrium. That is, the player reaches the Nash equilibrium when the player's mutual information reaches the channel capacity. Next, we demonstrate the security and fairness of the proposed protocol in detail.
\subsection{Nash Equilibrium Proof}
\textbf{Theorem 3}. There is a pair of strategies $(s_1^*,s_2^*)$ in the game that makes the protocol reach the Nash equilibrium state.

In a standard-form game $G=\{P_1,\dots,P_n;S_1,\dots,S_n;u_1,\dots,u_n\}$ with $n$ players, if $n$ is bounded, and for each $i$, $S_i$ is a non-empty convex set in Euclidean space, the utility function $u_i$ is continuous with respect to $s_i$, and the pair of $s_i$ is quasi-convex Concave, the game has a Nash equilibrium.

\textbf{Proof.} In standard form game $G=\{P_1,\dots,P_n;S_1,\dots,S_n;u_1,\dots,u_n\}$, there are two players, i.e. $n$ is bounded. simultaneously, for strategy $S_1=\{0\leq x_i\leq 1:1\leq i\leq n, x_1+x_2+\dots+x_n=1\}$, it is an $n-1$-dimensional subcube in an $n$-dimensional closed cube with side length 1, which is a non-empty convex set in Euclidean space. Similarly, for strategy $S_2=\{0\leq y_i\leq 1:1\leq i\leq m, y_1+y_2+\dots+y_n=1\}$, it is an $m-1$ dimensional subcube in an $m$-dimensional closed cube of side length 1,which is a non-empty convex set in Euclidean space. In addition, according to the concave-convexity theorem of mutual information, it can be known that when the channel $p(x|y)$ is fixed, $u_1$ is continuous to $s_1(s_1\in S_1)$, and is a concave function to $s_1$. Likewise, under the condition that the channel $p(x|y)$ is fixed, $u_2$ is continuous with respect to $s_2(s_2\in S_2)$ and a concave function with respect to $s_2$.

Since all the conditions in the game  $G=\{P_1,\dots,P_n;S_1,\dots,S_n;u_1,\dots,u_n\}$ are satisfied, that is, there is a pure strategy Nash equilibrium in the constructed standard game  $G=\{P_1,\dots,P_n;S_1,\dots,S_n;u_1,\dots,u_n\}$. In other words, there is a pair of pure strategies $s_1^*=(p_1^*,p_2^*,...,p_n^*)$ and $s_2^*=(q_1^*,q_2^*,...,q_n^*)$ in a standard game, which correspond to a pair of input and output variables $X^*$ and $Y^*$. When $ P(X^*=i)=p_i^*,1 \leq i \leq n, p_1^*+p_2^*+,...,+p_n^*=1$ and $P(Y^*=j)=q_j^*,1\leq j \leq m, q_1^*+q_2^*+,...,+q_m^*=1,$ it makes any given $s_2\in S_2$ must have $u_1(s_1^*,s_2)\geq u_1(s_1,s_2)$ and any given $s_1\in S_1$ and $u_2(s_1,s_2^*)\geq u_2(s_1,s_2)$ must be established at the same time. Furthermore, according to the definition of channel capacity, we know of channel $C=u_1(s_1^*,s_2^*)=u_2(s_1^*,s_2^*)$ of channel $p(y\vert x)$, it means that the $P_1$ and $P_2$ reach the Nash equilibrium during the game, which is equal to the channel capacity when the player $P_1$ is the sender of the channel and the player $P_2$ is the receiver of the channels, i.e. $u_i=C=\underset{p(x)}{max}\ \vert I(\textbf{a},X;\textbf{b},Y)\vert=\underset{q(y)}{max}\vert I(\textbf{b},Y;\textbf{a},X)\vert$.

\begin{table}[h]
		\centering
	\caption{Direction entropy and utility of players.}
	\label{tab:Direction entropy and utility of players.}
	\begin{tabular}{ccccc}
		\toprule
		Serial&Strategy & Direction vector&Direction entropy&Utility\\
		\toprule
		(1)&\{(C,C)\}&\{$a_i=1,b_i=1$\}&\{$H(dir)_t,H(dir)_t$\}&\{$\vert I(\textbf{a},X;\textbf{b},Y)_{t}\vert,\vert I(\textbf{b},Y;\textbf{a},X)_{t}\vert$ \}\\
		(2)&\{(C,D)\}&\{$a_i=1,b_i=-1$)&\{$H(dir)_t,H(dir)_{t-1}$\}&\{$\vert I(\textbf{a},X;\textbf{b},Y)_{t}\vert,\vert I(\textbf{b},Y;\textbf{a},X)_{t-1} \vert$\}\\
		(3)&\{(D,C)\}& \{$a_i=-1,b_i=1$)&\{$H(dir)_{t-1},H(dir)_t$\}& \{$\vert I(\textbf{a},X;\textbf{b},Y)_{t-1}\vert,\vert I(\textbf{b},Y;\textbf{a},X)_{t}\vert$\}\\
		(4)&\{(D,D)\}&\{$a_i=-1,b_i=-1$\} &\{$H(dir)_{t-1},H(dir)_{t-1}$\}&\{$\vert I(\textbf{a},X;\textbf{b},Y)_{t-1}\vert,\vert I(\textbf{b},Y;\textbf{a},X)_{t-1}\vert$\}\\
		\bottomrule
	\end{tabular}
\end{table}

\subsection{Security proof}
\textbf{Theorem 4} The rational and secure two-party computation protocol based on direction entropy is secure.

\textbf{Proof}: The protocol performs security proofs in terms of safe abort and correct delivery of outputs.

Safe abort: Since both parties do not know the specific value of $t$, the two parties can only finalize the secret after $n$ interactions. Therefore, neither party can terminate the agreement at will, and if it is terminated at will, they will not be able to obtain the secret.

Correct delivery of output: (1) Suppose there is an adversary $A$ who can forge a random sub-secret $s^{'}_i$ and send it to the participant during the secret exchange process, then the player calculates $Pr[I(S;s^{'}_i)=I(S;{s_i}_{i\in[n]})]\leq negl$ for the sub-secret $s^{'}_i$, that is, the player cannot distinguish the random value from the sub-secret. The probability is negligible.

(2) Assuming that there is an adversary $A$ who can forge a duplicate sub-secret s during the secret exchange process and send it to the player, then the player calculates $Pr[I(S;s^{''}_i)-I(S;{s_i})]\leq negl$ for the sub-secret  $s^{''}_i$, that is, the probability that the player cannot distinguish the duplicate sub-secret  $s^{''}_i$ is negligible.

(3) After the protocol is executed, all participating parties can integrate their own secrets. The adversary's advantage is $Adv^{\pi}_A(\widetilde{s},y)=Pr[y|y\gets\underset{[x]}{\bigoplus}\widetilde{s}]\leq negl$, where s represents the sub-secret forged by the adversary, that is, the probability that the adversary can integrate the secret when the external adversary $A$ forges the sub-secret input can be ignored.

\subsection{Proof of fairness}\label{subsec3}
\textbf{Theorem 5.} The rational and secure two-party computation protocol based on direction entropy is fair.

\textbf{Proof.} Since both players reach the Nash equilibrium when the channel capacity is reached. At the same time, their utility is maximized when they reach the channel capacity. In the secret exchange phase, players judge the amount of information transmitted according to their own channel capacity. In each round of information exchange, players judge the amount of information transferred based on their channel capacity. Specifically, we obtain the utility function as shown in Table \ref{tab:Direction entropy and utility of players.}. to get the following three situations:

Case 1: When $M(q)>N(p)$, it means that the probability of player $P_2$ chooses $C$ is greater than the probability of player $P_1$ chooses  $C$ during the $t\in\left\{1,2,\cdots,n\right\}$ round of information interaction. Overall, the mutual information of the player $P_2$ is greater than the mutual information of the player $P_1$, that is, the utility of the player $P_2$ is greater than the utility of the player $P_1$. At this time, player $P_2$ regards the information delivered by player $P_1$ as incorrect information, and $P_2$ terminates the protocol.

Case 2: When $M(q)<N(p)$, it means that the probability of player $P_2$ chooses $C$ is less than the probability of player $P_1$ chooses $C$ during the $t\in\left\{1,2,\cdots,n\right\}$ round of information interaction. Overall, the mutual information of the player $P_1$ is greater than the mutual information of the player $P_2$, that is, the utility of the player $P_1$ is greater than the utility of the player $P_2$. At this time, player $P_1$ regards the information delivered by player $P_2$ as incorrect information, and $P_1$ terminates the protocol.

Case 3: When $M(q)=N(p)$, it means that the probability of player $P_1$ chooses $C$ is equal to the probability of player $P_2$ chooses $D$ during the $t\in\left\{1,2,\cdots,n\right\}$ round of information interaction. The mutual information obtained by players $P_1$ and $P_2$ is equal. In other words, the utility of players $P_1$ and $P_2$ are equal. When the player chooses to cooperate with a probability close to 1, the player maximizes the utility and the protocol reaches the Nash equilibrium. In short, if the player chooses $C$ with a probability close to 1, the player successfully exchanges information and the protocol reaches the optimal state.

\section{Simulations and results}
\label{sec: Simulations and results}
In this section, we first utilize the MATLAB software to simulate and analyze the player's utility. Subsequently, we also analyze the utility function of players when they choose cooperative strategies with different probabilities. We then analyze the performance of the protocol.

\subsection{Experiments}
\begin{figure}[h!]
	\centering
	\includegraphics[width=0.6\textwidth]{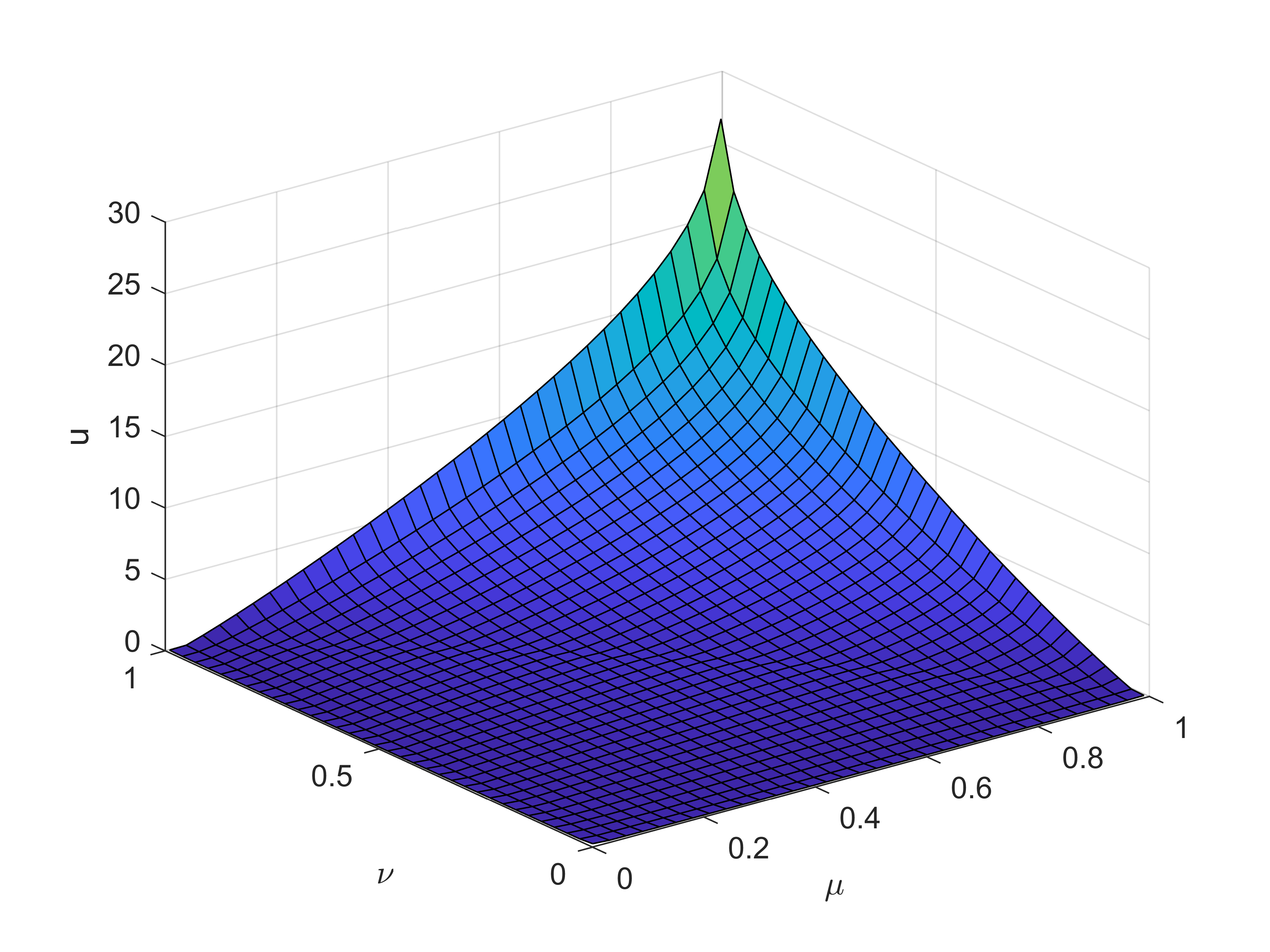}
	\caption{The utility changes of player}
	\label{fig:The utility changes of player}       % Give a unique label
\end{figure}

In figure \ref{fig:The utility changes of player},we shows a graph of a player's utility as a function of the probability of cooperation. Where $\mu$ represents the probability of player $P_1$ choosing a cooperative strategy, $\nu$ represents the probability of player $P_2$ choosing a cooperative strategy, and $u$ represents the player's utility. It can be seen from the figure that the utility $u$ of the player gradually increases with the increase of $\mu$ and $\nu$, that is, the greater the probability that the player chooses the cooperative strategy, the more utility he obtains. In addition, when the probability of players $P_1$ and $P_2$ choosing a cooperative strategy is 1,  their utility is maximized.

\begin{figure}[h!]
	\centering
	\includegraphics[width=0.6\textwidth]{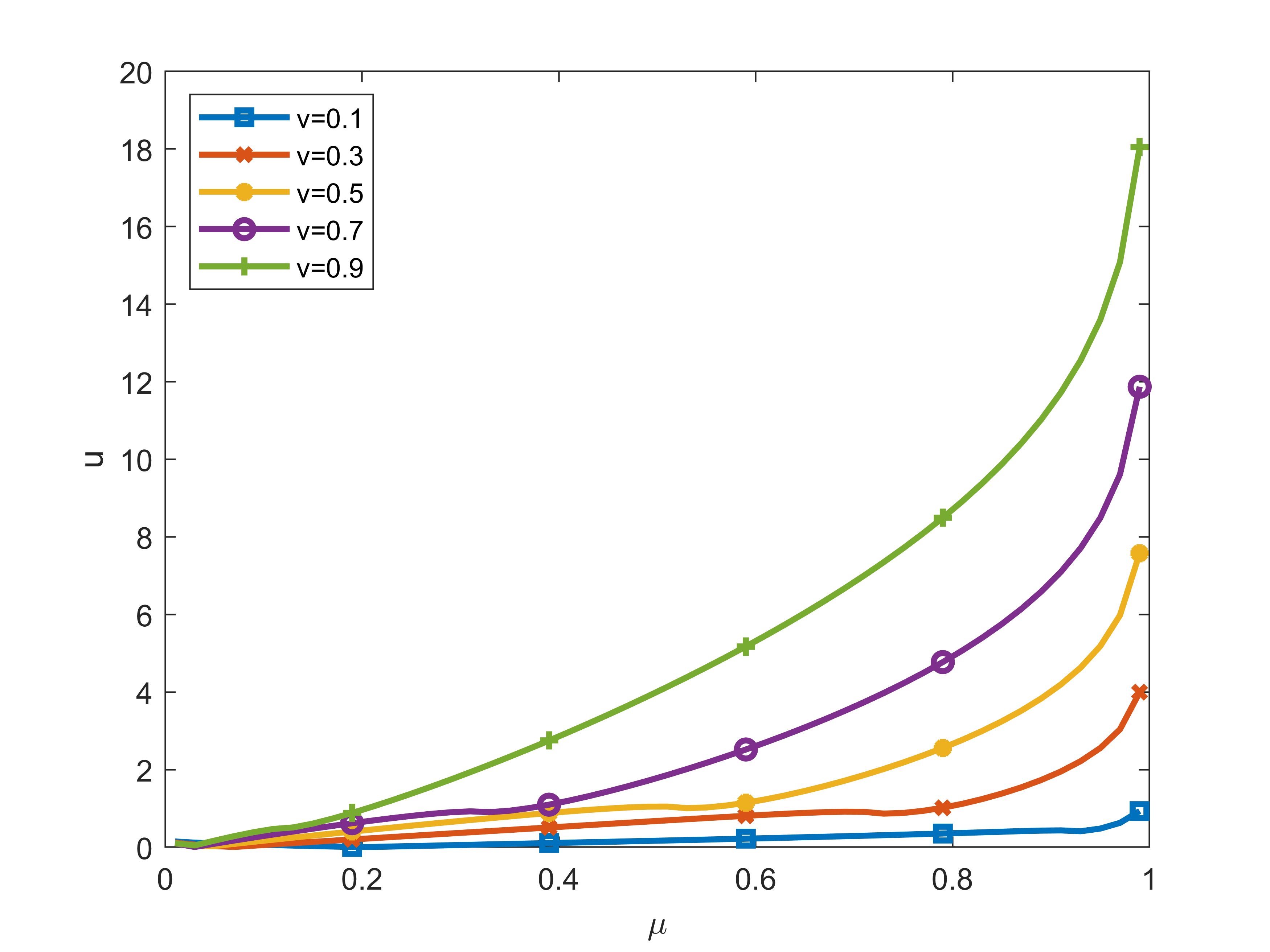}
	\caption{The utility changes of player $P_1$}
	\label{fig:The utility changes of player $P_1$}       % Give a unique label
\end{figure}

In figure \ref{fig:The utility changes of player $P_1$}, we show a graph of the player's probability of choosing a cooperative strategy as a function of the utility gained. Among them, we use $\mu$ to denote the probability of player $P_1$ choosing a cooperative strategy, and $u$ to denote the utility of player $P_1$. Specifically, when the $\nu$ values are 0.1, 0.3, 0.5, 0.7 and 0.9 respectively, it can be seen from the figure that the probability of player $P_1$ choosing a cooperative strategy is positively correlated with the utility function. At the same time, the probability of player $P_2$ choosing a cooperative strategy also has a positive correlation with the maximum utility value of player $P_1$. That is, player $P_2$ determines the maximum utility value of player $P_1$, and vice versa.

\subsection{Performance analysis}
\label{Performance analysis}
The table \ref{tab:The comparison of protocol.} shows the comparison of the RSTC in the proposed scheme and the existing RSMC. Comparing from the privacy, fairness and utility. Here, "Yes" satisfies the performance and "No" dissatisfies the performance.
\begin{table}[h!]
	\centering
	\caption{The comparison of protocol.}
	\label{tab:The comparison of protocol.}
	\begin{tabular}{ccccc}
		\toprule
		Serial &Literature& Security &Fairness &Utility \\
		\toprule
		(1) &Wang et al. \cite{20}&Yes&Yes&Entropy\\
		(2) &Zhang et al. \cite{21}&Safety within threshold&Yes&Deposit\\
		(3) &	Ah-Fat et al. \cite{22}&Based on TTP&Based on TTP&/\\
		(4) &Our protocol &Yes& Yes& Mutual information\\
		\bottomrule
	\end{tabular}
\end{table}

We leverage game theory to propose the direction entropy-based solution. Our protocol realize the privacy and fairness of RSTC. Based on TTP, the protocol proposed by Ah-Fat et al. \cite{22} also achieves fairness, but due to the existence of TTP, this will lead to leakage of the privacy of the agreement. Wang et al. \cite{20} construct the player's utility function through information entropy, and use belief to solve the player's fairness problem. Zhang et al. \cite{21} pre-deposit deposits and construct agreements to make players choose honest strategies to achieve the fairness of the agreement. Different from these protocols, our protocol constructs a utility function through mutual information, and is based on directional entropy to ensure the fairness of the protocol. In addition, both players reach a Nash equilibrium when the channel capacity is reached.

\section{Conclusion}
\label{sec:Conclusion}
In this study, we proposed a rational secure two-party computing protocol based on direction entropy, which has a game nature. Although our protocol does not set a deposit or a specific utility function as the existing rational protocol, our protocol still achieves its fairness, as well as guaranteeing the correctness of the final output. We adopt the direction vector of direction entropy to quantify the player's strategy from different dimensions. Specifically, the player chooses a cooperation strategy, and its direction vector is positive, which means that the information delivered is positive. Conversely, it means that the information passed is negative. Furthermore, according to the player's behavioral preferences, we define the player's utility based on mutual information. The proof and analysis results show that players gain utility when they comply with the protocol. What's more, when the player chooses a cooperation strategy with a probability close to 1, its mutual information reaches the maximum, which is equal to the channel capacity. 

In future work, one of our directions is to apply our protocol to a specific scenario like in \cite{37}, which will bring more valuable effects. Furthermore, in this paper, based on the direction entropy, we propose a rational secure two-party computation protocol. However, in real scenarios there are usually multiple players. Therefore, another direction is to extend our protocol to one-to-many or even many-to-many protocols.

\section*{Acknowledgment}%基金
This article is supported by the National Key Project of China(No.2020YFB1005700); Natural Science Foundation under Grant(Grant Number:61962009); National Key Research and Development Program of China(No.2021YFA1000600); National Key Research and Development Program of Guangdong Province under Grant(2020B0101090002); Major Scientific and Technological Special Project of Guizhou Province under Grant(20183001,20193003); Foundation of Guangxi Key Laboratory of Cryptography and Information Security(GCIS202118); Science and Technology Support Plan of Guizhou Province([2020]2Y011) and the Key Research and Development Program of Guangzhou(No.202103050003).

\bibliographystyle{cas-model2-names}

 %l\cite{yang2019containment}

\bibliography{ref2.bib}

\end{document}